\documentclass[aps,prl,twocolumn,superscriptaddress,groupedaddress]{revtex4}  
\usepackage{graphicx}  
\usepackage{dcolumn}   
\usepackage{bm}        
\usepackage{amssymb}   
\usepackage[latin9]{inputenc}
\usepackage{amsmath}
\usepackage{subfigure}
\usepackage{amssymb}
\usepackage{graphicx}
\usepackage{amsfonts}%
\usepackage{dcolumn}
\usepackage{bm}
\usepackage{hyperref}
\usepackage{nccmath}
\usepackage{lipsum}
		{\end{pmatrix}\end{medsize}}%

\begin{document}
\title{Tunable SSH model in ferromagnetic systems}
\author{Chi-Ho Cheung}
\email{f98245017@ntu.edu.tw}
\author{Jinyu Zou}
\email{jinyuzou@hust.edu.cn}
\affiliation{Wuhan National High Magnetic Field Center $\&$ School of Physics, Huazhong University of Science and Technology, Wuhan 430074, China}

	\date{\today}

	\begin{abstract}
	It is well known that the topology of Su-Schrieffer-Heeger (SSH) model, which belongs to AIII symmetry class, is protected by chiral symmetry. In this article, instead of chiral symmetry, we constrain the bulk Hamiltonian by a magnetic point group symmetry, which can be generated by a unitary symmetry and an anti-unitary symmetry. Under these symmetries, a four-band model can be block-diagonalized into two 2-band models and each 2-band model is analogous to an SSH model. As the two 2-band models are individual, we call the four-band model double independent Su-Schrieffer-Heeger (DISSH) model. Interestingly, since the symmetry requirements of DISSH model can be fulfilled in ferromagnetic systems, the discovery in this article extends SSH model into ferromagnetic systems. Furthermore, we present an example of DISSH model with a set of reasonable parameters to show that it is possible to manipulate the topological phase of DISSH model by tuning the magnetic moment in experiment.

	\end{abstract}
	
	\maketitle
	
	\section{I. Introduction}
In the academic world, since David Thouless, Duncan Haldane, Michael Kosterlitz and others introduced topology into the field of solid-state physics in a few decades ago[1-3], it has caused great repercussions. Topology is a mathematical discipline that mainly studies the properties of a geometric object that are preserved under continuous deformations. It includes studying the topological classification of the mapping between two closed manifolds. A solid with periodic symmetry has a first Brillouin zone and a set of occupied state wave functions, both of which are continuous and closed (sometimes occupied state wave functions can be represented by occupied Bloch vectors). The mapping between the first Brillouin zone and the occupied state wave functions or occupied Bloch vectors can be regarded as a mapping between two closed manifolds. Thus, topology can be applied to the field of solid-state physics.

Topology is widely used in the field of solid-state physics. In the early stage, it mainly focused on the application of electronic systems, including quantum anomalous Hall effect (QAHE) in magnetic insulators[3-6], quantum spin Hall effect (QSHE) in non-magnetic insulators[7-11], and topological crystal insulator (TCI) in magnetic and non-magnetic insulators[12-18], Weyl semimetal[19-26], Weyl nodal line semimetal[27, 28], Dirac semimetal[29-31], triple point topological metal[32], sixfold and eightfold degenerate semimetal[34]. Chern number and winding number[35] are widely used in these topological materials to indicate the nontrivial phase.

The Su-Schrieffer-Heeger (SSH) model[36, 37] is the most intuitive and understandable among all kinds of solid-state physical models related to topology. The original SSH model is a one-dimensional two-band model. Under the restriction of chiral symmetry, there are only two components of Bloch vectors left in the two-band model. Therefore, the mapping between one-dimensional Brillouin zone and the closed trajectory swept by Bloch vector is described by the mapping from $S^1$ to $S^1$, which is topologically classified by $Z$ invariant. Therefore, the topological invariant of the SSH model can be obtained by computing the winding number. Although SSH model is the most intuitive topological solid-state physical model from the theoretical point of view, it is not easy to be realized in experiments. The main reasons are that it is difficult to produce a one-dimensional material in experiments, and the chiral symmetry does not exist in any magnetic space group. In recent years, due to the improvement of experimental techniques for realizing one-dimensional or quasi-one-dimensional systems, finding other symmetries to replace chiral symmetry, expanding the application of topology and other factors, scientists have been keen to realize SSH model in various systems such as phonon system[38], optical system[39, 40], circuit system[41], paramagnetic or antiferromagnetic electronic system[42].

The original SSH model is constrained by chiral symmetry. In this article, instead of chiral symmetry, we constrain the bulk Hamiltonian by a magnetic point group symmetry which can be generated by a unitary symmetry and an anti-unitary symmetry. Since the unitary symmetry can divide the Hilbert space of a four-band model into two 2-dimensional subspaces and the anti-unitary symmetry further limits the Bloch vectors of the two subspaces, the topological classification of Bloch vector trajectories in these two subspaces is the same as the one in SSH model. Because the magnetic space group of some ferromagnetic systems is able to meet the symmetry requirements, the discovery in this paper extends SSH model into ferromagnetic systems. Furthermore, we use a tight-binding(TB) model with a set of reasonable parameters to show that, there is a possibility of manipulating topological phase transition by tuning the magnitude of magnetic moment in experiment.

\section{II. Symmetry constraints of DISSH model}
If a symmetry group has two generators, a unitary operator $U$ and an anti-unitary operator $\mathcal{V}$, satisfying the following equation:
\begin{equation}\label{eq1}
	\begin{split}
		&U^2=-1, \\
		&\mathcal{V}^2=1, \\
		&U\mathcal{V}+\mathcal{V}U=0.
	\end{split}
\end{equation}
Then the pair of $U$ and $\mathcal{V}$ as shown below,
\begin{equation}\label{eq2}
	\begin{split}
		&U=i\tau_z \sigma_z, \\
		&\mathcal{V}=-i\tau_0\sigma_z K,
	\end{split}
\end{equation}
is the solution of eq.\ref{eq1}, where $K$ is the complex conjugate operator, $\tau_{x,y,z}$ and $\sigma_{x,y,z}$ are the Pauli matrices, $\tau_0$ and $\sigma_0$ are the $2 \times 2$ identity matrices.

Any Hermitian four-band Hamiltonian can be expanded with 16 $\Gamma$ matrices. Among them, only six $\Gamma$ matrices are commutative with $U$ and $\mathcal{V}$ simultaneously. The six $\Gamma$ matrices are $\tau_0 \sigma_0, \tau_z \sigma_z, \tau_z \sigma_0, \tau_0 \sigma_z, \tau_x \sigma_y$ and $\tau_y \sigma_x$. The unitary operator $U$ can divide the Hilbert space of the $4\times 4$ Hamiltonian into two 2-dimensional subspaces, one is corresponding to eigenvalue $i$ of $U$ and the other one is corresponding to eigenvalue $-i$ of $U$. For the $2\times 2$ Hamiltonian of $i$ subspace, $\tau_0 \sigma_0, \tau_z \sigma_z, \tau_z \sigma_0, \tau_0 \sigma_z, \tau_x \sigma_y$ and $\tau_y \sigma_x$ play the role of $\sigma_0, \sigma_0, \sigma_z, \sigma_z, \sigma_y$ and $\sigma_y$ respectively. For the $2\times 2$ Hamiltonian of $-i$ subspace, $\tau_0 \sigma_0, \tau_z \sigma_z, \tau_z \sigma_0, \tau_0 \sigma_z, \tau_x \sigma_y$ and $\tau_y \sigma_x$ play the role of $\sigma_0, -\sigma_0, \sigma_z, -\sigma_z, -\sigma_y$ and $\sigma_y$ respectively. Hence, the Hamiltonian of $\pm i$ subspace can be written in the following form:
\begin{equation}\label{eq3}
	H_{\pm i}(k) = d_{\pm i,0}\sigma_0 + d_{\pm i,z}\sigma_z + d_{\pm i,y}\sigma_y.
\end{equation}
Since $d_{\pm i,x}\sigma_x$ is forbidden by symmetries, and $d_{\pm i,0}\sigma_0$ does not affect the position of the Wannier Center[43], Eq.\ref{eq3} implies that if the Hamiltonian on a one-dimensional line in the momentum space commutes with $U$ and $\mathcal{V}$ simultaneously, the winding number on this one-dimensional line is restricted to 0 or $\pi$, just likes the winding number of SSH model.

\begin{figure*}
	\centering
	\includegraphics[width=1\textwidth]{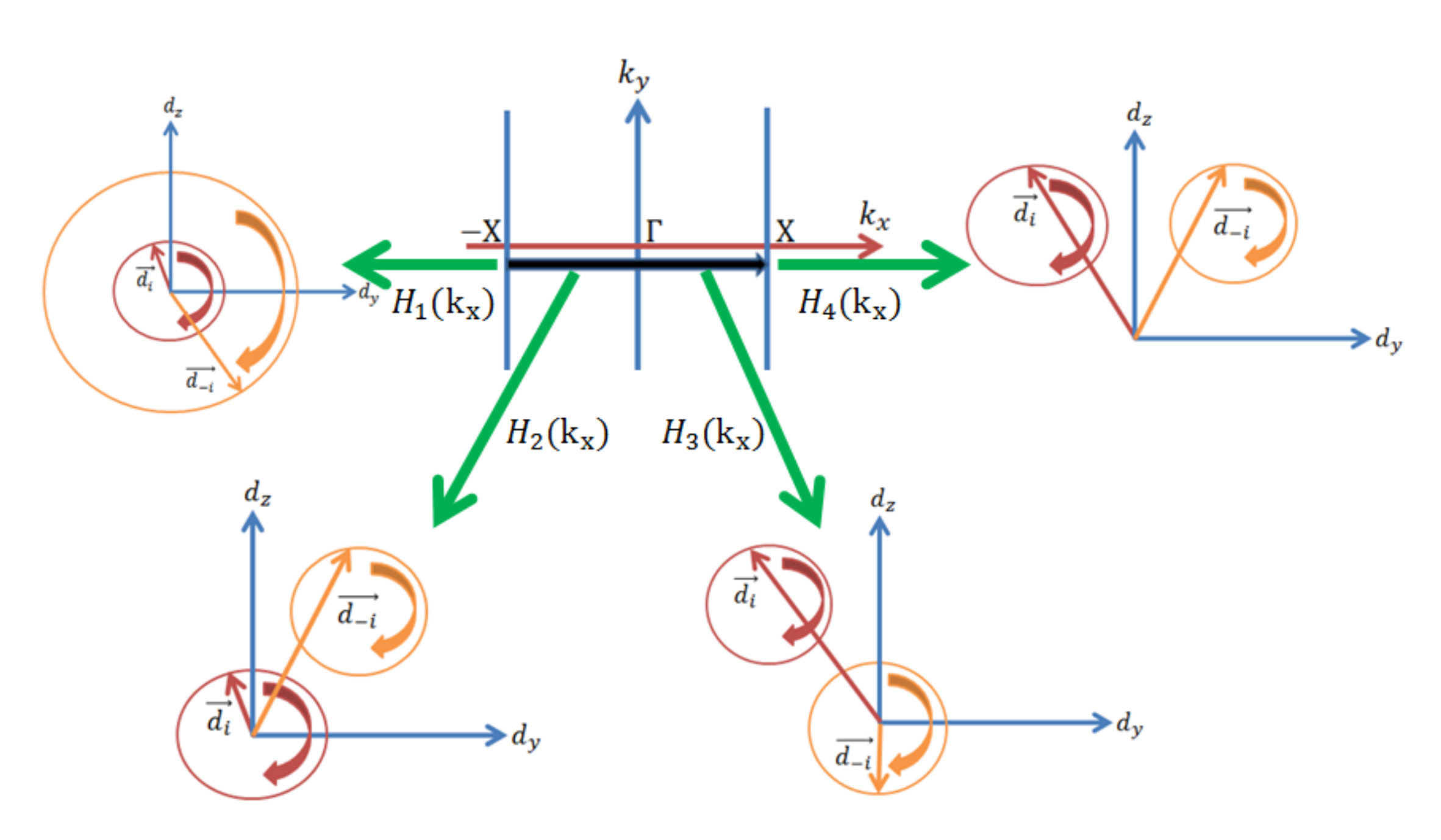}\\
	\caption{A variety of topologically unequal trajectories of the DISSH model on the line of $k_y=0$. $\vec{d}_{\pm i}$ represents the Bloch vector of $\pm i$ subspace. $H_1 (k_x)$, $H_2 (k_x)$, $H_3 (k_x)$ and $H_4 (k_x)$ are four topologically unequal bulk Hamiltonians. }\label{Fig1}
\end{figure*}

For example, if a two-dimensional symmorphic system lying on the xy plane has $M_z$ and $M_xT$ ($T$ is time-reversal symmetry), then its Hamiltonian has to commute with $M_z$ and $M_xT$ on the line of $k_y=0$. Furthermore, if the two-dimensional system is insulated on the $k_y=0$ line, then the momentum space on this line has a variety of topologically unequal trajectories as it is mapped to Bloch vector space by a variety of Hamiltonians(as shown in Fig.\ref{Fig1}). On the line of $k_y=0$, both $\vec{d}_{+i}$ and $\vec{d}_{-i}$ are restricted to $d_y-d_z$ plane, but the trajectories of $\vec{d}_{+i}$ and $\vec{d}_{-i}$ can be different. Therefore, the subspace with eigenvalue $i$ of $M_z$ and the subspace with eigenvalue $-i$ of $M_z$ form two independent SSH models. We call these two independent SSH models separated by two eigenvalues of a unitary operator as double independent Su-Schrieffer-Heeger (DISSH) model.

In a broad sense, it is not difficult to realize DISSH model in electronic systems while considering spin orbit coupling (SOC). Firstly, there is no special requirement in dimension, DISSH can be realized in one-, two- and three-dimensional systems. Secondly, it is possible to realize DISSH model in two- and three-dimensional metal systems, because DISSH model only needs two or three-dimensional systems to be insulated on a high symmetry path whose little group contains $U$ and $\mathcal{V}$. Thirdly, although all the examples in this article are symmorphic systems, non-symmorphic systems may be able to realize DISSH model while the fractional translation operations attached to $U$ and $\mathcal{V}$ do not cause $U$ and $\mathcal{V}$ violating the restriction of eq.\ref{eq1}. Fourthly, systems with higher symmetry can realize DISSH model while symmetries other than $U$ and $\mathcal{V}$ neither cause degeneracy on the high symmetry path whose little group contains $U$ and $\mathcal{V}$ nor eliminate the second or third term in Eq.\ref{eq3}. Fifthly, systems contain $U$ and $\mathcal{V}$ are very common because $U$ and $\mathcal{V}$ can be caused by mirror symmetries or $C_2$ rotational symmetries. Both $C_2$ rotational symmetries and mirror symmetries are common symmetries in single-crystal systems.

\begin{figure}[htbp]
	\centering
	\includegraphics[width=8.5cm]{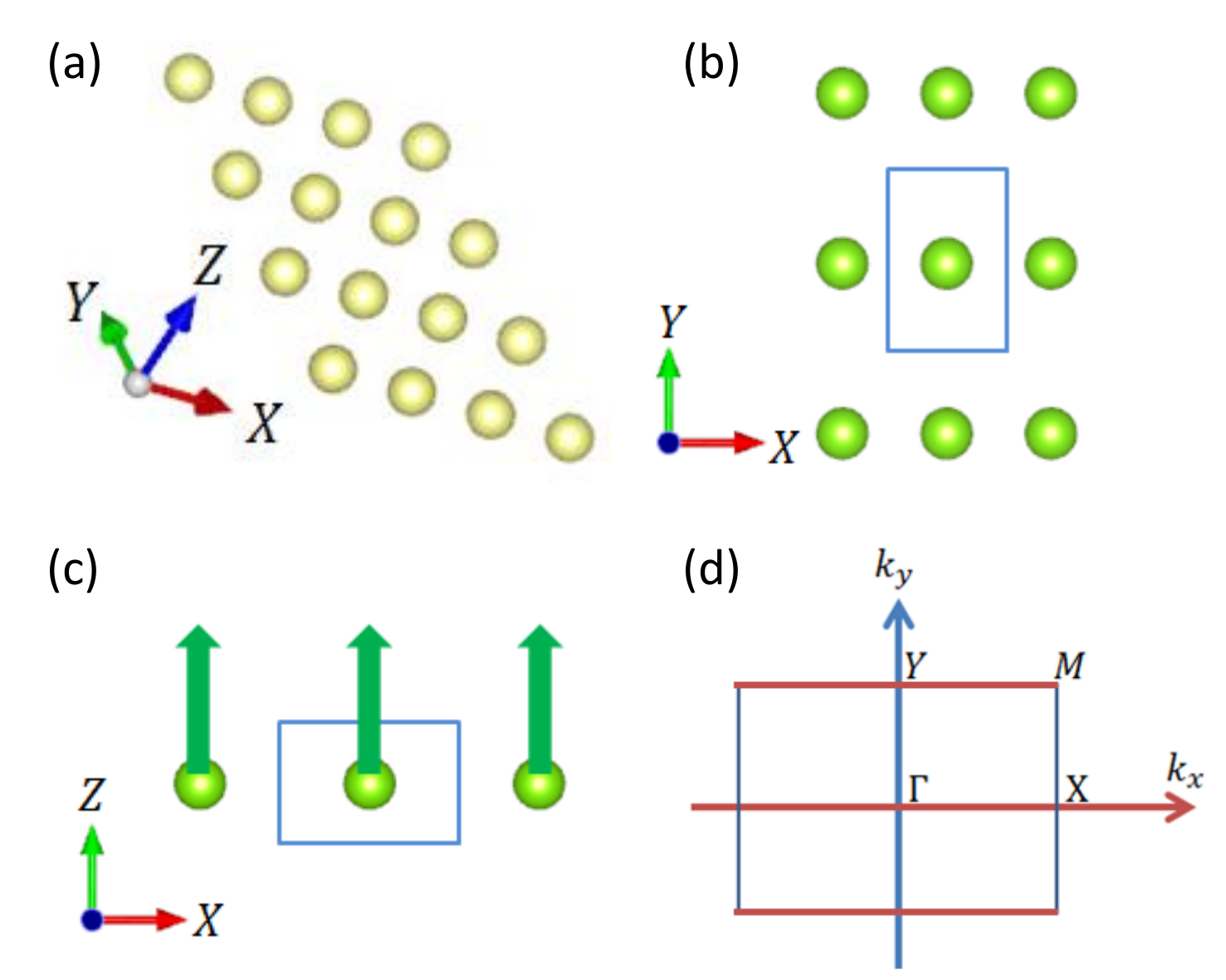}\\
	\caption{(a)-(c) show the 2D system with $Pmmm$ crystal structure. The green arrows in (c) represent the direction of magnetic moment of each atom. (d) shows the first Brillouin zone of the 2D system. The red lines in (d) represent two DISSH models. }\label{Fig2}
\end{figure}

\section{III. Tunable DISSH model in ferromagnetic system}
In this section, we use a tight-binding(TB) model of a ferromagnetic system to show that DISSH model can exist in ferromagnetic systems. Besides, the TB model with different sets of reasonable parameters shows that it is possible to manipulate topological phase transition by tuning the magnitude of magnetic moment. It is worth emphasizing that the TB model discussed in this section is just an example. The more general discussion for DISSH model is presented in the previous section.

Assume a 2D system with $Pmmm$ crystal symmetry contains one atom in its primitive unit cell and each atom owns a magnetic moment which points to $z$ direction(as shown in Fig.\ref{Fig2}). In nonmagnetic phase, the generators of this system are $M_z$, $M_x$, $T$ and Inversion symmetry. Furthermore, assume the only atom in each primitive unit cell contributes four orbitals which are $|s_{\uparrow}\rangle, |s_{\downarrow}\rangle, |p_{z\uparrow}\rangle$ and $|p_{z\downarrow}\rangle$. The representation of $M_z$, $M_x$, $T$ and Inversion symmetry in these four bases are:
\begin{equation}\label{eq4}
	\begin{split}
		M_z &= i\tau_z \sigma_z, \\
		M_x &= i\tau_0 \sigma_x, \\
		T   &= i\tau_0 \sigma_y K, \\
		I   &= \tau_z \sigma_0,
	\end{split}
\end{equation}
where $I$ is the Inversion symmetry.

\begin{figure*}[htbp]
	\centering
	\includegraphics[width=1\textwidth]{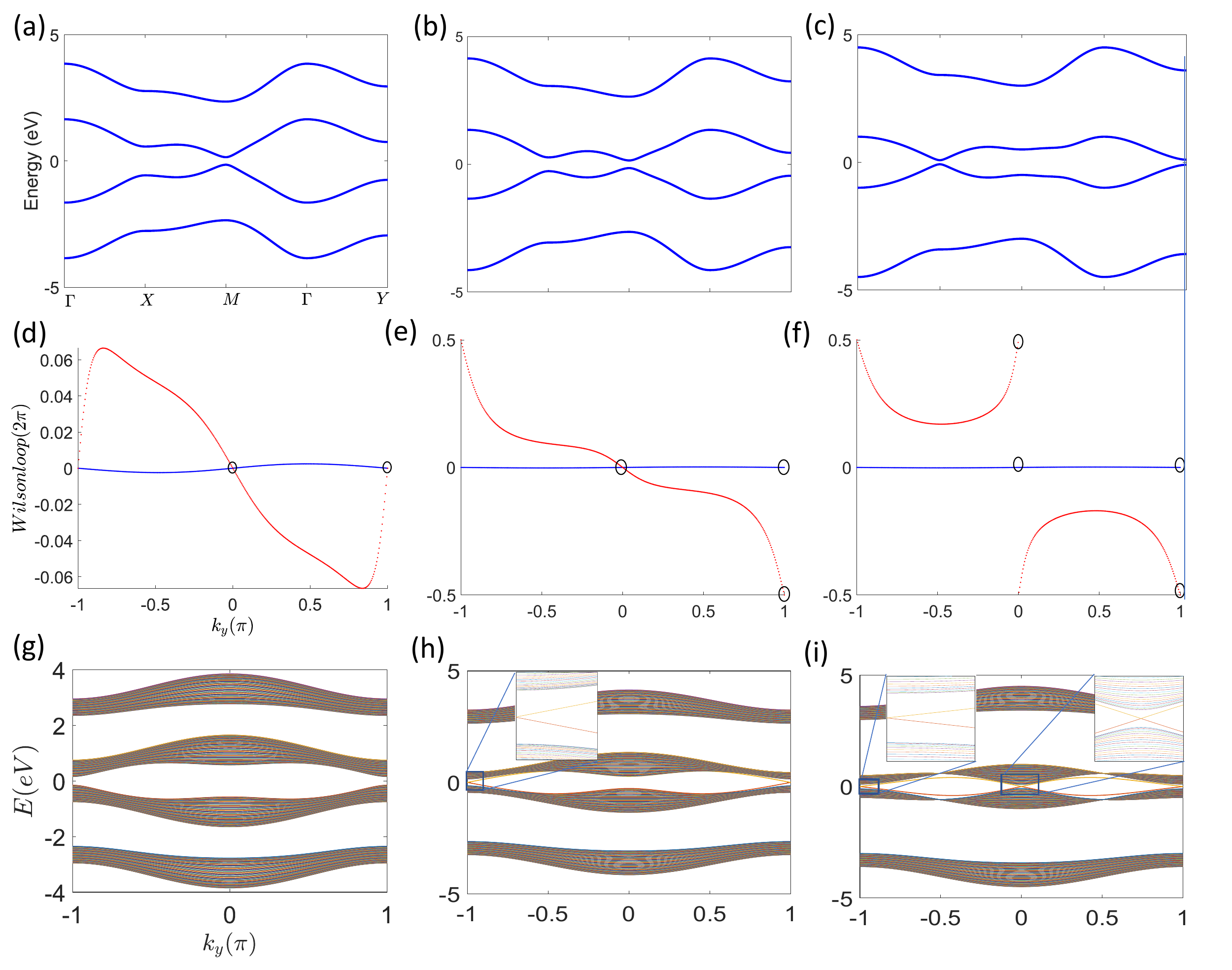}\\
	\caption{(a), (b) and (c) are the bulk band structures with $Z_0=1.1 eV$, $1.4 eV$ and $1.75 eV$ respectively. The bulk Hamiltonian is given in Eq.\ref{eq5} while all the parameters in Eq.\ref{eq5} are listed in Table.\ref{Table1}. (d), (e) and (f) are the Wannier Centers corresponding to the bulk band structures in (a), (b) and (c) respectively. The position of Wannier Centers of DISSH models is marked in black circle. (g), (h) and (i) are the ribbon band structures corresponding to the bulk band structures in (a), (b) and (c) in respectively. The insets in (h) and (i) show the zero-energy modes of DISSH models. }\label{Fig3}
\end{figure*}

For bulk band analysis, calculating winding number and plotting the energy band of edge states, a TB model of bulk Hamiltonian is needed. By method of invariants and Eq.\ref{eq4}, a TB model for the 2D ferromagnetic system in Fig.\ref{Fig2} is constructed as follows:
\begin{equation}\label{eq5}
	\begin{split}
		H(k) &= p_1 \tau_z \sigma_0 - p_2 \sin{k_x}\tau_x \sigma_y + p_3 \cos{k_x}\tau_z \sigma_0 \\
		&+ p_4 \cos{k_y}\tau_z\sigma_0 - p_5 \sin{k_y}\tau_x\sigma_x \\
		&- 2p_6 \sin{k_x}\cos{k_y}\tau_x\sigma_y + 2p_7 \cos{k_x}\cos{k_y}\tau_z\sigma_0 \\
		&+ 2p_8 \cos{k_x}\sin{k_y}\tau_x\sigma_x + Z_0 \tau_0\sigma_z.
	\end{split}
\end{equation}

\begin{table}[tbp]  
	\centering
	\caption{The parameters of the TB model. The first row lists the name of the parameters in Eq.\ref{eq5}. The second row gives the value(s) of each parameter in the unit of $eV$. The parameter $Z_0$ represents the magnitude of Zeeman term which varies as the magnitude of the magnetic moment varies.}\label{Table1}
		\begin{tabular}{c|c|c|c|c|c|c|c|c}
			\hline
			\hline
			$p_1$ & $p_2$ & $p_3$ & $p_4$ & $p_5$ & $p_6$ & $p_7$ & $p_8$ & $Z_0$ \\
			\hline
			1.88 & 0.45 & 0.42 & 0.33 & 0.37 & 0.05 & 0.06 & 0.07 & 1.1 or 1.4 or 1.75 \\
			\hline
		\end{tabular}
\end{table}

The parameters $p_{1-8}$ and $Z_0$ are listed in Table.\ref{Table1}. Up to the second nearest neighbor, all terms preserving $D_{2h}$, $T$ and particle-hole symmetry are included in Eq.\ref{eq5}. The last term in Eq.\ref{eq5} is the Zeeman term which is induced by the magnetic moment. Only the last term breaks $T$ and $M_x$, but it preserves inversion symmetry, $M_z$ and $M_xT$. For keeping Eq.\ref{eq5} simple, inversion symmetry and particle-hole symmetry are added to the equation, besides, except the Zeeman term, all the other terms which could be induced by the magnetic moment are ignored. However, according to the analysis in the previous section, any terms preserving $M_z$ and $M_xT$ and not closing the bulk gap on the red lines in Fig.\ref{Fig2}(d) cannot change the topological properties of DISSH model. Therefore, it is safe to write down the bulk Hamiltonian as simple as Eq.\ref{eq5}.

\begin{figure}[htbp]
	\centering
	\includegraphics[width=8.5cm]{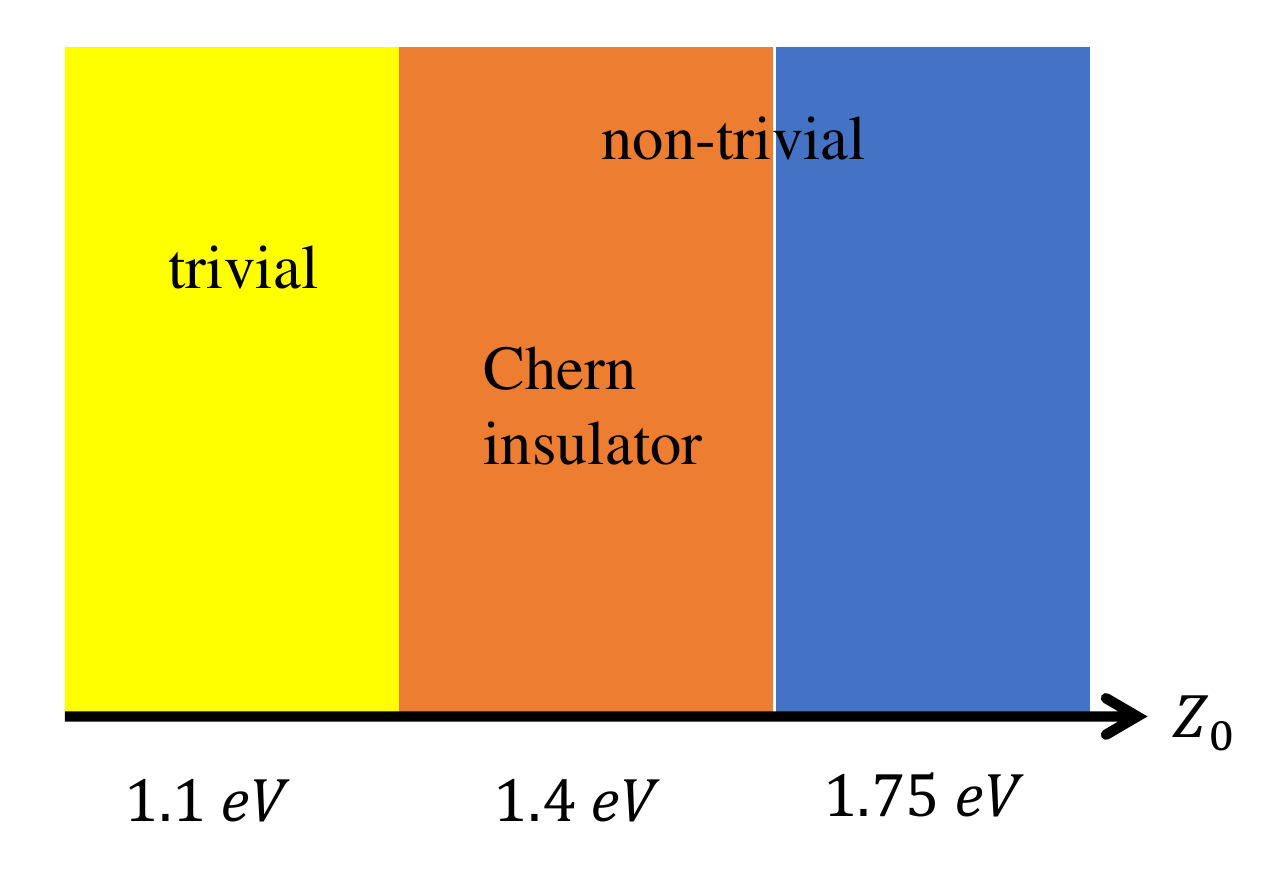}\\
	\caption{The phase diagram of the TB model in Eq.\ref{eq5} with various $Z_0$. All the parameters in Eq.\ref{eq5} are listed in Table.\ref{Table1}. Yellow color represents that the two DISSH models on the line of $k_y=0$ and $k_y=\pi$ are trivial. Red color and blue color represent that at least one of the two DISSH models is non-trivial while red color represents that the TB model in Eq.\ref{eq5} is a model of Chern insulator.}\label{Fig4}
\end{figure}

Since the magnitude of magnetic moment is tunable in experiment and the effect of changing the magnetic moment on the Hamiltonian is mainly reflected in the Zeeman term $Z_0$, we choose to manipulate the topological phase of DISSH model by tuning the Zeeman term. Firstly, the DISSH models at the line of $k_y=0$ and of $k_y=\pi$ are trivial when $Z_0=1.1 eV$, such that, there are no zero-energy modes in ribbon band structures. Secondly, a band inversion occurs at $M$ point in the subspace with eigenvalue $i$ of $M_z$ (hereafter $i$ subspace means the subspace with eigenvalue $i$ of $ M_{z} $) when $Z_0$ continuously changes from $1.1 eV$ to $1.4 eV$, such that, the DISSH model at the line of $k_y=\pi$ becomes non-trivial while the DISSH model at the line of $k_y=0$ remains trivial. At this state, the 2D system is a Chern insulator with Chern number equal to 1. Hence, there are edge states of Chern insulator and zero-energy modes at $k_y=\pi$ appear in ribbon band structures. Thirdly, a band inversion occurs at $X$ point in the $i$ subspace when $Z_0$ continuously changes from $1.4 eV$ to $1.75 eV$, such that, the DISSH models at the line of $k_y=0$ becomes non-trivial too. At this state, although the 2D system is no longer a Chern insulator, zero-energy modes of the two DISSH models still appear at $k_y=0$ and $k_y=\pi$ in ribbon band structures. The bulk band structures, Wilson loops and ribbon band structures with different $Z_0$ are shown in Fig.\ref{Fig3}. The phase diagram with various $Z_0$ is shown in Fig.\ref{Fig4}. Fig.\ref{Fig3} and Fig.\ref{Fig4} are plotted with the parameters listed in Table.\ref{Table1}.

One more thing is worth notice. Even if all the SOC terms in Eq.\ref{eq5} are one tenth of the original SOC terms, Zeeman term $Z_0$ with $1.1eV$, $1.4eV$ and $1.75eV$ still correspond to the three different topological phases discussed above. Hence, the topological phases shown in Fig.\ref{Fig3} and Fig.\ref{Fig4} are not sensitive to the SOC terms in Eq.\ref{eq5}.

\section{IV. Conclusion}
We constrain the bulk Hamiltonian by a magnetic point group which can be generated by a unitary symmetry $U$ and an anti-unitary symmetry $\mathcal{V}$. $U$ can be mirror symmetry or $C_2$ rotational symmetry while $\mathcal{V}$ can be mirror symmetry multiplies $T$ or $C_2$ rotational symmetry multiplies $T$, both mirror symmetry and $C_2$ rotational symmetry are very common in single-crystal systems.

Since $U$ can divide the Hilbert space of a four-band model into two 2-dimensional subspaces and $\mathcal{V}$ further limits the Bloch vectors of the two subspaces, the topological classification of Bloch vector trajectories in these two subspaces is the same as the one in SSH model. In addition, because the trajectories of the two 2-band models can be different, we call the four band model double independent Su-Schrieffer-Heeger (DISSH) model. As the symmetry requirements allow the DISSH model to exist in ferromagnetic system, the discovery disclosed in this paper extends SSH model into ferromagnetic systems.

Furthermore, we use a tight-binding(TB) model with a set of reasonable parameters to show that it is possible to manipulate the topological phase of DISSH model by tuning the magnitude of magnetic moment in experiment.

\quad

\end{document}